%% file: zznlogg_off.tex
\documentclass[12pt,a4paper]{article}
\pdfoutput=1
\usepackage[utf8]{inputenc}
\usepackage{epsf,amsmath,amssymb,graphicx,dcolumn}
\usepackage{caption}
\usepackage{subcaption}
\usepackage{scalefnt,ulem,pstricks}
\usepackage{booktabs,multirow,tabularx}
\usepackage[colorlinks=true,allcolors={blue!70!black}]{hyperref}
\usepackage{cleveref}
\usepackage{color}
\usepackage{rotating}
\usepackage{cancel}
\usepackage{microtype}
\usepackage[titletoc,title]{appendix}

\usepackage[numbers,sort&compress]{natbib}
\usepackage{amssymb}
\usepackage{bm, amsbsy}
\usepackage{xspace}

\newcommand{\AZone}{\mathcal{A}_{gg,{\rm bkg}}^{(1\text{-loop})}}
\newcommand{\AHone}{\mathcal{A}_{gg,H}^{(1\text{-loop})}}
\newcommand{\AZtwo}{\mathcal{A}_{gg,{\rm bkg}}^{(2\text{-loop})}}

\newcommand{\AHtwo}{\mathcal{A}_{gg,H}^{(2\text{-loop})}}

\newcommand{\AZonemassl}{\mathcal{A}_{gg,{\rm bkg}}^{(1\text{-loop,massless})}}
\newcommand{\AHonemassl}{\mathcal{A}_{gg,H}^{(1\text{-loop,massless})}}
\newcommand{\AZtwomassl}{\mathcal{A}_{gg,{\rm bkg}}^{(2\text{-loop,massless})}}
\newcommand{\AHtwomassl}{\mathcal{A}_{gg,H}^{(2\text{-loop,massless})}}

\input{defs}

\begin{document} 
\hypersetup{pageanchor=false}
\begin{titlepage}
\renewcommand{\thefootnote}{\fnsymbol{footnote}}
\begin{flushright}
     MPP-2021-15\\
     ZU-TH 06/21
     \end{flushright}
\par \vspace{10mm}

\begin{center}
{\Large \bf Four lepton production in gluon fusion:\\[5pt] off-shell Higgs effects in NLO QCD}
\end{center}

\par \vspace{2mm}
\begin{center}
  {\bf Massimiliano Grazzini}$^{(a)}$, {\bf Stefan Kallweit}$^{(b)}$,\\[0.2cm]
  {\bf Marius Wiesemann}$^{(c)}$ and {\bf Jeong Yeon Yook}$^{(a)}$

$^{(a)}$ Physik-Institut, Universit\"at Z\"urich, 8057 Z\"urich, Switzerland 

$^{(b)}$ Dipartimento di Fisica, Universit\`a degli Studi di Milano-Bicocca and
INFN, Sezione di Milano-Bicocca, 20126, Milan, Italy

$^{(c)}$ Max-Planck-Institut f\"ur Physik, F\"ohringer Ring 6, 80805 M\"unchen, Germany

\end{center}

\begin{center} {\bf Abstract} \end{center}\vspace{-1cm}
\begin{quote}
\pretolerance 10000

We consider the production of four charged leptons in hadron collisions and compute the next-to-leading order (NLO) QCD corrections to the loop-induced gluon fusion contribution by
consistently accounting for the Higgs boson signal, its corresponding background and their interference.
The contribution from heavy-quark loops is exactly included in the calculation except for the two-loop $gg\to ZZ\to 4 \ell$ continuum diagrams, for which the unknown heavy-quark effects are approximated through a reweighting procedure.
Our calculation is combined with the next-to-next-to-leading order QCD and NLO electroweak 
corrections to the $q{\bar q}\to 4\ell$ process, 
including all partonic channels and consistently accounting for spin correlations and off-shell effects.
The computation is implemented in the {\sc Matrix} framework and allows us to separately study the Higgs boson signal, the background and the interference contributions, whose knowledge can be used to constrain the Higgs boson width through off-shell measurements.
Our state-of-the-art predictions for the invariant-mass distribution of the four leptons are in good agreement with recent ATLAS data.
\end{quote}

\vspace*{\fill}
\begin{flushleft}
February 2021
\end{flushleft}
\end{titlepage}
\hypersetup{pageanchor=true}

The observation of a scalar resonance by the ATLAS and CMS experiments \cite{Aad:2012tfa,Chatrchyan:2012ufa} at the Large Hadron
Collider (LHC) in 2012 marked a milestone towards our understanding of the mechanism of electroweak (EW) symmetry breaking. Studies of its spin, parity and couplings are in good agreement with the hypothesis that the new particle is the Standard Model (SM) Higgs boson.
Those studies are mainly focused on on-shell Higgs production. Although for
a SM Higgs boson of mass $m_H=125$ GeV the expected width is $\Gamma_H\sim 4$ MeV, with $\Gamma_H/m_H\sim 3\times 10^{-5}$,
it is well known that off-shell Higgs production has a substantial rate \cite{Kauer:2012hd}.
Furthermore, the interference between Higgs-mediated $gg\to H\to VV$ production and continuum $gg \to VV$ production 
is strong and destructive in the high invariant-mass region, which is required to preserve unitarity for the scattering amplitudes of massive fermions and gauge bosons.

Off-shell Higgs boson production plays an important role for the determination of the Higgs boson decay width at the LHC.
Indeed, the limited resolution of current detectors ($\sim 1 \, \GeV{}$) prevents a direct measurement of $\Gamma_H$.
Fortunately, the ratio of the off-shell to the on-shell cross sections can be used \cite{Caola:2013yja, Campbell:2013una, Campbell:2013wga, Ellis:2014yca} to set stringent constraints on $\Gamma_H$~\cite{Khachatryan:2014iha,Aad:2015xua,Khachatryan:2015mma,Khachatryan:2016ctc,Aaboud:2018puo,Sirunyan:2019twz}.
For instance, in a scenario in which the Higgs couplings and width are rescaled from their SM values such that the on-shell cross section remains unchanged \cite{Caola:2013yja}, the off-shell signal rate would scale with $\Gamma_H/\Gamma_H^{\rm SM}$, while
the interference would scale with $\sqrt{\Gamma_H/\Gamma_H^{\rm SM}}$. Therefore, in order to constrain $\Gamma_H$ through off-shell measurements, it is essential
to have a precise description of four-lepton production and a separation between the Higgs boson signal, the background and their interference.
In the following, we focus on the $H\to ZZ$ channel and present state-of-art predictions for the three contributions, including the leptonic decays of the $Z$ bosons with off-shell effects and spin correlations.

At leading order (\LO{}) in the strong coupling $\alpha_S$,  continuum $ZZ$ production occurs via quark annihilation. The next-to-leading-order (\NLO{}) QCD prediction for on-shell $ZZ$ production has been known for almost three decades \cite{Mele:1990bq, Ohnemus:1990za} as well as with leptonic decays~\cite{Ohnemus:1994ff, Dixon:1999di, Campbell:1999ah, Dixon:1998py}. The
\NLO{} electroweak (EW) corrections were first computed for on-shell $Z$ bosons~\cite{Accomando:2004de, Bierweiler:2013dja, Baglio:2013toa}, and off-shell effects were included in \citeres{Biedermann:2016yvs, Biedermann:2016lvg}. In addition, \NLO{} QCD+EW predictions have been calculated for the $2\ell 2\nu$~\cite{Kallweit:2017khh} and $2\ell 2\ell'$~\cite{Chiesa:2018lcs} final states. The loop-induced gluon fusion channel enters the computation at $\mathcal{O}(\alpha_S^2)$. Its \LO{} on-shell contribution was first calculated in \citeres{vanderBij:1988fb, Dicus:1987dj}, followed by the inclusion of leptonic decays~\cite{Matsuura:1991pj, Zecher:1994kb, Binoth:2008pr, Campbell:2011cu, Kauer:2013qba, Cascioli:2013gfa, Campbell:2013una, Ellis:2014yca, Kauer:2015dma}. The NLO QCD corrections to the loop-induced gluon fusion channel were calculated in \citere{Caola:2015psa,Caola:2016trd, Alioli:2016xab} for the gluon--gluon~(\gg{}) partonic channel using the two-loop amplitudes of \citeres{Caola:2015ila,vonManteuffel:2015msa}, and the full NLO QCD corrections 
including the quark--gluon (\qg{}) initial states were presented in \citere{Grazzini:2018owa}.
NLO corrections to the Higgs signal--background interference have been considered in \citeres{Campbell:2016ivq,Caola:2016trd}.
The full next-to-next-to-leading order (NNLO) QCD corrections (to the quark initiated production) have been obtained both for on-shell $Z$ bosons~\cite{Cascioli:2014yka, Heinrich:2017bvg} and for the fully differential 
leptonic final states~\cite{Grazzini:2015hta, Kallweit:2018nyv}, using the $\qqx\to VV'$ two-loop amplitudes of \citeres{Gehrmann:2014bfa,Caola:2014iua,Gehrmann:2015ora}. Recently, also the combination of \NNLO{} QCD and \NLO{} EW predictions has been achieved \cite{Kallweit:2019zez} using \Matrix{}~\cite{Grazzini:2017mhc} and \OpenLoops{}~\cite{Cascioli:2011va, Buccioni:2017yxi, Buccioni:2019sur}. Finally, by combining the NNLO quark-initiated cross section and the NLO gluon-initiated cross section \citere{Grazzini:2018owa} presented also 
approximate N$^{3}$LO predictions (labelled as $\nNNLO{}$). Their combination with \NLO{} EW predictions analogous to the results presented in \citere{Grazzini:2020stb} for $W^+W^-$ production can be achieved with \Matrix{} as well.

In this Letter, we present new results for four-lepton production at the LHC. We start from the calculation of \citere{Grazzini:2018owa} and perform a more sophisticated treatment of the heavy-quark mass 
effects in the two-loop amplitudes of the \gg{} initiated
production. As before, both \gg{} and \qg{}  partonic channels are considered in the NLO QCD corrections.
In addition, we separately study the $gg\to H\to 4\ell$ signal cross section, the four-lepton continuum background as well as their interference, which are the relevant theoretical
ingredients to constrain $\Gamma_H$.
We also provide state-of-the-art numerical predictions by combining \nNNLO{} QCD and \NLO{} EW corrections, and we compare them 
with recent ATLAS data at $13$\,TeV \cite{Aaboud:2019lxo}. 

\begin{figure}[t]
\begin{center}
\begin{tabular}{ccc}
\includegraphics[width=.24\textwidth]{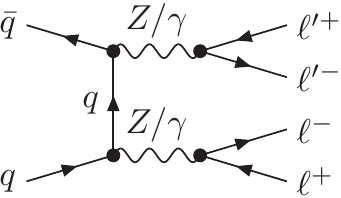}& \quad\quad\quad\quad &
\includegraphics[width=.24\textwidth]{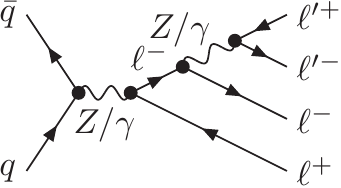}\\[1ex]
(a) & \quad & (b) 
\end{tabular}
\end{center}
\caption[]{\label{fig:qq}{Sample Feynman diagrams for the production of four charged leptons in
 the quark annihilation channel at LO: (a) $t$-channel and (b) Drell--Yan-like topologies.}}
\vspace*{4ex}
\begin{center}
\begin{tabular}{ccc}
$\AZone{}$ $\widehat{=}$ \parbox{0.35\textwidth}{\includegraphics[width=.35\textwidth]{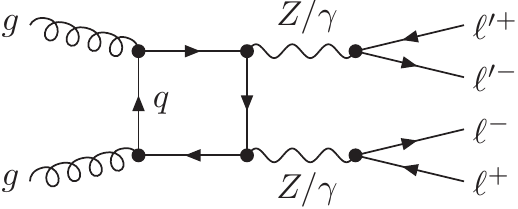}}& &
$\AHone{}$ $\widehat{=}$ \parbox{0.35\textwidth}{\includegraphics[width=.35\textwidth]{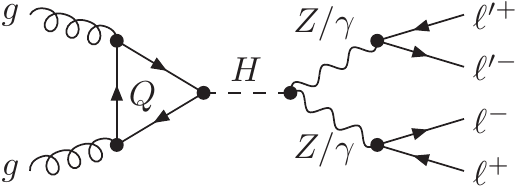}}\\[6ex]
(a) & \quad & (b)\\[4ex]
$\AZtwo{}$ $\widehat{=}$ \parbox{0.35\textwidth}{\includegraphics[width=.35\textwidth]{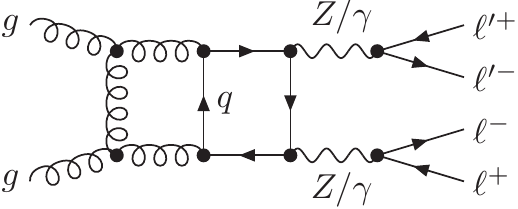}}& &
$\AHtwo{}$ $\widehat{=}$ \parbox{0.35\textwidth}{\includegraphics[width=.35\textwidth]{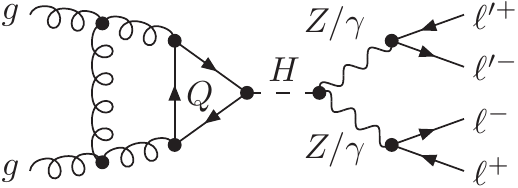}}\\[6ex]
(c) & \quad & (d)
\end{tabular}
\end{center}
\caption[]{\label{fig:gg}{Loop-induced diagrams for the production of four charged leptons in the gluon fusion channel at one-loop level (a,b) and at two-loop level (c,d),
separately for the continuum $ZZ$ contributions (a,c) and the Higgs-mediated contributions (b,d). Here, $q\in\{d,u,s,c,b,t\}$ denotes all quarks, whereas $Q\in\{b,t\}$ denotes the quarks treated as massive in this Letter.}}
\end{figure}

\begin{figure}[t]
\begin{center}
\begin{tabular}{c}
\includegraphics[width=.6\textwidth]{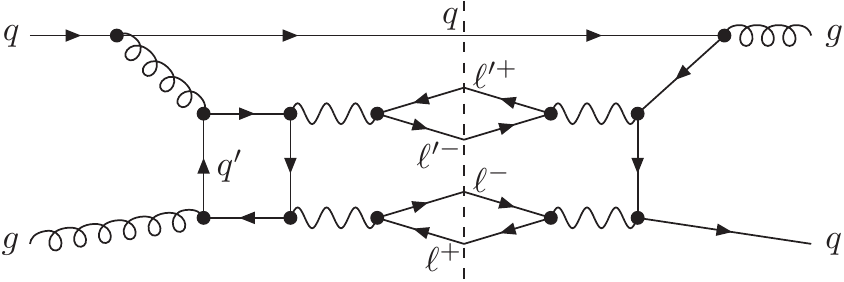}\\
\end{tabular}
\end{center}
\caption[]{\label{fig:mix}{Example of \NNLO{} interference between quark annihilation and loop-induced gluon fusion production mechanisms.}}
\end{figure}

We consider the four-lepton production process
\begin{align}
pp \rightarrow \ell^+\ell^-\,\ell'^+\ell'^-+X\,,\nonumber
\end{align}
both in the same-flavour channel ($\ell' = \ell$) and in the different-flavour channel ($\ell' \neq \ell$), where $\ell, \ell' \in  {e, \mu}$. At the LO, this process is driven by quark annihilation. 
Representative Born-level diagrams are shown in \fig{fig:qq}\,(a) and (b) for $t$-channel and Drell--Yan-like topologies, respectively.
Starting from NNLO in QCD the contribution from loop-induced diagrams driven by gluon fusion have to be taken into account. \fig{fig:gg}\,(a) and (b)
show sample diagrams of the four-lepton continuum and the Higgs-mediated production, respectively.
Although those are suppressed by two powers of the strong coupling, their enhancement by the large gluon flux renders them quantitatively important. 
As a result, the inclusion of NLO corrections to the loop-induced gluon fusion contribution is crucial for precision physics of four-lepton production. 
Note, however, that the quark annihilation and loop-induced gluon fusion processes cannot be treated as being completely independent.
Indeed, they mix already at NNLO in QCD, and \fig{fig:mix} illustrates an example. Such contributions have to be included, and 
the interference renders the distinction between the two production mechanisms cumbersome. 
As in \citeres{Grazzini:2018owa, Grazzini:2020stb} we obtain a partial N$^{3}$LO result, labelled as $\nNNLO{}$ in the following, by combining the NNLO QCD predictions
with NLO QCD corrections to the loop-induced gluon fusion contribution, including all partonic channels, but considering only diagrams with purely fermionic loops. 
Any other N$^{3}$LO contributions cannot be included consistently at present, and thus they are not considered in our calculation. Nevertheless, those contributions can be expected to be sub-dominant 
with respect to the corrections we include at $\nNNLO{}$.

In this Letter, take a few decisive steps in order to advance our calculation in \citere{Grazzini:2018owa}.
In particular, we improve the description of the Higgs signal and signal--background interference by evaluating exactly all the contributions
  that do not depend on the continuum $gg\to \ell^+\ell^-\,\ell'^+\ell'^-$ two-loop helicity amplitude (see below).
We perform the calculation by explicitly separating the Higgs boson signal, the four-lepton continuum background, and 
their interference. Those are the underlying theoretical ingredients the experimental analyses require in order to constrain the Higgs boson width.
Finally, as done for $W^+W^-$ production in \citere{Grazzini:2020stb} we supplement our $\nNNLO{}$ predictions with NLO corrections in the EW coupling expansion, using the implementation presented 
in \citere{Kallweit:2019zez}.

Our calculation includes the complete dependence on heavy-quark masses in all contributions, but in the two-loop helicity amplitudes of $q\bar{q}\to \ell^+\ell^-\,\ell'^+\ell'^-$
and $gg\to\ell^+\ell^-\,\ell'^+\ell'^-$, where they are unknown\footnote{Very recently,
the computation of the top-quark contribution to the two-loop on-shell $gg\to ZZ$ helicity amplitudes has been reported \cite{Agarwal:2020dye,Bronnum-Hansen:2021olh}.}.
For the quark annihilation process the contribution of closed fermion loops is relatively small, and heavy-quark effects
can be safely neglected at two-loop level. By contrast, such effects are important for the loop-induced gluon fusion process,
where they enter effectively at LO, i.e.\ $\mathcal{O}(\as^2)$, through 
one-loop diagrams, see \fig{fig:gg}\,(a) and (b). While at one-loop level the full mass dependence is known
and included throughout our calculation, at two-loop level, see \fig{fig:gg}\,(c) and (d),
the heavy-quark effects for the continuum amplitude in \fig{fig:gg}\,(c)  have not yet been computed.
As it is well known, the impact of heavy-quark loops is particularly relevant for the Higgs signal--background interference,
since in the high-mass region the off-shell Higgs boson decays to longitudinally polarised $Z$ bosons, which
in turn have a stronger coupling to heavy quarks.
Using an appropriate reweighting procedure the missing heavy-quark contributions can be approximated. In the 
following, we discuss in detail the approach used in \citere{Grazzini:2020stb} and the improvement pursued here.
To this end, we define the finite part (in the hard scheme \cite{Catani:2013tia}) of the one-loop and
two-loop amplitudes $\AZone{}$, $\AHone{}$ and $\AZtwo{}$, $\AHtwo{}$ separately for the four-lepton continuum background
and the Higgs-mediated contribution, as indicated in \fig{fig:gg} with one sample diagram for each amplitude.
The $\as$  expansion of the full $gg\to \ell^+\ell^-\,\ell'^+\ell'^-$ amplitude and its square can then be written as
\begin{align}
  \label{eq:amp}
  \mathcal{A}_{gg\to 4\ell} =&\; \as\,\left(\AZone{}+\AHone{}\right) + \as^2\,\left(\AZtwo{}+\AHtwo{}\right)+\mathcal{O}(\as^3)\,,\\
|\mathcal{A}_{gg\to 4\ell}|^2 =&\; \as^2\,\left(\AZone{}+\AHone{}\right)^* \,\left(\AZone{}+\AHone{}\right) \notag\\
&+ \as^3\times2\,{\rm Re}\left[\left(\AZone{}+\AHone{}\right)^*\,\left(\AZtwo{}+\AHtwo{}\right)\right]+\mathcal{O}(\as^4)\,.
\end{align}
  In fact, each of the amplitudes in \eqn{eq:amp} is known with its full heavy-quark mass dependence, except for $\AZtwo{}$, 
  which is known only for massless quark loops. In \citere{Grazzini:2020stb} we have computed the entire $\as^3$ contribution
  to the squared amplitude in the massless approximation, reweighted with the full mass dependence at one-loop,
\begin{align}\label{eq:approx1}
&\hspace*{-3em}\left(\AZone{}+\AHone{}\right)^*\,\left(\AZtwo{}+\AHtwo{}\right)\\
\approx&\; \frac{\left(\AZonemassl{}+\AHonemassl{}\right)^*\,\left(\AZtwomassl{}+\AHtwomassl{}\right)}{\left(\AZonemassl{}+\AHonemassl{}\right)^*\,\left(\AZonemassl{}+\AHonemassl{}\right)}\notag\\
&\times  \left(\AZone{}+\AHone{}\right)^*\,\left(\AZone{}+\AHone{}\right)\,,\notag
\end{align}
where $\AHonemassl=\AHtwomassl=0$. However, also the two-loop Higgs form factor relevant to compute $\AHtwo{}$ is known
including the full heavy-quark mass effects \cite{Spira:1995rr,Anastasiou:2006hc,Harlander:2005rq,Aglietti:2006tp}.
In particular, in the new implementation we use the explicit expression of \citere{Harlander:2019ioe} and
combine it with the $gg\to H \to \ell^+\ell^-\,\ell'^+\ell'^-$ one-loop amplitude,
taking care of the correct complex phases in the amplitude definition, to obtain the full result for $\AHtwo{}$.
In a second step, we apply a judicious reweighting procedure, using the full one-loop amplitudes to approximate
the mass effects in all contributions interfered with $\AZtwo{}$,
\begin{align}\label{eq:approx2}
&\hspace*{-3em}\left(\AZone{}+\AHone{}\right)^*\,\AZtwo{}\\
\approx&\ \frac{\left(\AZone{}+\AHone{}\right)^*\,\AZtwomassl{}}{\left(\AZone{}+\AHone{}\right)^*\,\AZonemassl{}}\times  \left(\AZone{}+\AHone{}\right)^*\,\AZone{}\,.\notag
\end{align}
Note that this reweighting procedure is implemented at the level of the squared/interfered amplitudes,
since this amounts to simply multiplying complex numbers, rather than at the amplitude level before summing over helicities,
which would be more involved. However, \eqn{eq:approx2} effectively corrects only for the missing quark-mass effects in $\AZtwo{}$.
It is clear that with this approximation we obtain a much better treatment of the heavy-quark effects,
especially of the Higgs contributions, than using \eqn{eq:approx1}. 
In fact, with the new implementation the Higgs signal does not include any approximation.
One part of the interference contribution is complete as well, while the other part includes the mass effects 
of the one-loop correction. Only the background contribution is treated essentially in the same approximation as in \citere{Grazzini:2020stb}.
However, given that also the NNLO $q\bar{q}$ cross section is part of the background to the Higgs signal, this approximation is subleading.
In particular, the Higgs interference contribution in our approach is a new result including all contributions known to date,
which will be useful for constraining the Higgs width.

Our calculation is performed within the \Matrix{} framework~\cite{Grazzini:2017mhc}.
With \Matrix{}, NNLO{} QCD predictions can be obtained for various colour-singlet processes at hadron
colliders~\cite{Grazzini:2013bna,Grazzini:2015nwa,Cascioli:2014yka,Grazzini:2015hta,Gehrmann:2014fva,Grazzini:2016ctr,Grazzini:2016swo,Grazzini:2017ckn,Kallweit:2018nyv,Kallweit:2020gcp}.%
\footnote{It was also used in the NNLO+NNLL computation of \citere{Grazzini:2015wpa}, in the NNLO+N$^3$LL computations of \citere{Kallweit:2020gva,Wiesemann:2020gbm},
  and in the NNLO+PS computations of \citeres{Re:2018vac,Monni:2019whf,Alioli:2019qzz,Monni:2020nks,Lombardi:2020wju}.} 
The core of \Matrix{} is the Monte Carlo program \Munich{}~\cite{munich} which contains a fully automated implementation of the
dipole subtraction method~\cite{Catani:1996jh,Catani:1996vz} and an efficient phase space integration. \NLO{} corrections can be obtained
using either dipole subtraction or $q_T$ subtraction~\cite{Catani:2007vq}, which provides a self-consistency check for our results.
All tree-level and and one-loop amplitudes can be evaluated with either \OpenLoops{}\,2~\cite{Cascioli:2011va,Buccioni:2017yxi,Buccioni:2019sur}
or \Recola{}\,2~\cite{Actis:2016mpe,Denner:2017wsf}, and the corresponding numerical results are in full agreement.
In case of \OpenLoops{}, we use dedicated squared amplitudes to separate the Higgs signal, background and interference contributions in the gluon fusion channel.
In case of \Recola{}, we exploit the \texttt{SM\_FERM\_YUK} model to select the order 
of the top and bottom Yukawa couplings.
With this model, our \Recola{}\,2 implementation allows us to separate the Higgs signal and background at the level of helicity amplitudes.
Also in the calculation of the two-loop corrections, following \eqn{eq:amp} and using the approximation in \eqn{eq:approx2},
we exploit \Recola{}\,2 with the \texttt{SM\_FERM\_YUK} model to select the relevant one-loop helicity amplitudes $\AZone{}$ and $\AHone{}$.
For the two-loop amplitudes,
we exploit the calculation of the massless helicity amplitudes of \citere{vonManteuffel:2015msa}
that are implemented in {\sc VVamp} \cite{hepforge:VVamp} to obtain $\AZtwomassl{}$,
and we apply the two-loop Higgs form factor including the full heavy-quark mass effects of \citere{Harlander:2019ioe}
to the one-loop helicity amplitude $\AHone{}$ from \Recola{}\,2 in order to compute $\AHtwo{}$. To obtain the NNLO corrections to the quark-initiated process 
we exploit the general implementation of the $q_T$ subtraction formalism~\cite{Catani:2007vq} within \Matrix{} and
rely on the two-loop $q\bar{q}\to 4\ell$ helicity amplitudes of \citere{Gehrmann:2015ora} that are also provided by {\sc VVamp} \cite{hepforge:VVamp}.

Our implementation of NLO QCD corrections to the loop-induced gluon fusion production with separation of Higgs signal, background and interference 
has been validated by comparing fiducial and differential cross sections to the results of \citere
{Caola:2016trd}. \citere{Caola:2016trd} presents an NLO calculation of the Higgs signal,
the continuum $ZZ$ background and their interference, considering the $e^+e^-\mu^+\mu^-$ channel.
The calculation of \citere{Caola:2016trd} is limited to the \gg{} partonic channel and includes the top-quark loops in the two-loop $gg\to ZZ$ amplitude through a large-$m_t$ expansion.
To the purpose of our comparison we exactly reproduce the setup of \citere{Caola:2016trd}, except for the treatment of the bottom quarks,
which \citere{Caola:2016trd} considers as massless in the background amplitudes and as massive in
the Higgs-mediated amplitudes. \Matrix{}, on the other hand, treats bottom quarks as either massless or massive particles throughout the calculation in a consistent manner.
At LO we find complete agreement with the results of \citere{Caola:2016trd}, and we have independently checked our results with the
parton level Monte Carlo program MCFM \cite{Campbell:1999ah, Campbell:2011bn, Campbell:2015qma}.
At NLO we are able to reproduce the results of Eq.~(4) of \citere{Caola:2016trd} to better than $1\%$ percent.
We also find reasonably good agreement with the four-lepton invariant mass distributions reported in Fig.~6 of \citere{Caola:2016trd}.
Considering the different treatment of the bottom quarks, and the different approximation used for the top-quark contributions, we regard this agreement fully satisfactory.

\input{tables/Cuts_ATLAS.tex}

We now present predictions for $pp\rightarrow \ell^+\ell^-\ell'^+\ell'^-$ production at \mbox{$\sqrt{s} = 13\,\TeV$}.
The two lepton pairs $\ell^+\ell^-$ and $\ell'^+\ell'^-$ may have the same ($\ell=\ell'$)
or different ($\ell\neq\ell'$) flavours with $\ell, \ell' \in \{e,\mu\}$.
We use the selection cuts adopted in the ATLAS analysis of \citere{Aaboud:2019lxo}, summarized in \tab{tab:ZZcuts2}.
The three leading leptons must have transverse momenta $p_{T,l_1}$, $p_{T,l_2}$ and $p_{T,l_3}$ larger than 20, 15, and 10 GeV, respectively.
The fourth lepton is required to have $p_T> 7 (5)$ GeV for electrons (muons).
The electron and muon pseudorapidities must fulfil $|\eta_e|<2.47$ and $|\eta_\mu|<2.7$, respectively.
For each event, the lepton pair with an invariant mass $m_{12}$
closest to the $Z$ boson mass is required to have
$m_{12}$ in the range $50$ GeV $< m_{12} <$ $106$ GeV.
The remaining pair is referred to as the secondary pair, with mass $m_{34}$, and it must fulfil $f(m_{4\ell})<m_{34}<115$ GeV,
where the function $f(m_{4\ell})$ is \cite{Aaboud:2019lxo}
\begin{equation}
f(\mfourl) = \left\{ \begin{array}{ll} 5~\GeV, & \text{ for } \mfourl < 100~\GeV  \\  5~\GeV  + 0.7 \times \left(\mfourl - \
100~\GeV \right),  & \text{ for } 100~\GeV < \mfourl < 110~\GeV \\ 12~\GeV , & \text{ for } 110~\GeV <\mfourl < 140~\GeV   \\ 12~\GeV  + 0.76 \times \left(\mfourl - 140~\GeV\right),  & \text{ for } 140~\GeV  < \mfourl < 190~\GeV  \\ 50~\GeV ,  &\text{ for } \mfourl > 190~\GeV  \\  \end{array}\
\right..
\end{equation}
This selection strategy is tailored to preserve a good acceptance for low $\mfourl$ values,
but to suppress events with leptonic $\tau$ decays at higher $\mfourl$.
Leptons with different (same) flavours are separated by $\Delta R>0.2(0.1)$.
The invariant mass of each same-flavour opposite-sign lepton pair is required to be larger than $5$ GeV.
Finally, an invariant-mass range of 70 GeV $< m_{4\ell}<$ 1200 GeV is imposed on the four-lepton system.

For the electroweak parameters we use the $G_\mu$ scheme and set $\alpha=\sqrt{2}\,G_\mu m_W^2(1-m_W^2/m_Z^2)/\pi$.
The EW mixing angle is computed as $\cos\theta_W^2=(m_W^2-i\Gamma_W\,m_W)/(m_Z^2-i\Gamma_Z\,m_Z)$, and
the complex-mass scheme~\cite{Denner:2005fg} is used. The EW inputs are set to the PDG~\cite{Patrignani:2016xqp} values:
$G_F = 1.16639\times 10^{-5}$\,GeV$^{-2}$, $m_W=80.385$\,GeV, $\Gamma_W=2.0854$\,GeV,
$m_Z = 91.1876$\,GeV, $\Gamma_Z=2.4952$\,GeV, $m_H = 125$\,GeV, and $\Gamma_H = 0.00407$.
The on-shell top-quark and bottom-quark masses are set to $m_t = 173.2$\,GeV and $m_b = 4.5$\,GeV, respectively,
with $\Gamma_t=1.44262$\,GeV. Apart from the virtual two-loop contributions,
the full dependence on massive top and bottom quarks is taken into account everywhere in the computation,
and the four-flavour scheme with $N_f = 4$ massless quark flavours is used.
We employ the corresponding {\tt NNPDF31\_nnlo\_as\_0118\_luxqed\_nf\_4} \cite{Bertone:2017bme} PDF set
with $\as(m_Z) = 0.118$ at \LO{}, \NLO{}, and \NNLO{}.  The renormalization and factorization scales
are set dynamically to half of the four-lepton invariant mass, $\mu_R = \mu_F = m_{4\ell}/2$,
and scale uncertainties are estimated through customary 7-point scale variations with
the constraint $0.5 \leq \mu_R/\mu_F \leq 2$.

\input{tables/XS_ZZ_NNPDF31luxqedNf4.tex}

We start the presentation of our results in \tab{tab:XS_ZZ_2} with the fiducial cross sections
corresponding to the selection cuts in \tab{tab:ZZcuts2}.
We use the following notation: $q\bar{q}$NNLO refers to the NNLO result for
the $q\bar{q}$-initiated process, see \fig{fig:qq}, 
without the loop-induced gluon fusion
contribution; $gg$LO and $gg$NLO refer to the loop-induced 
gluon fusion contribution, see \fig{fig:gg}, at $\mathcal{O}(\as^2)$ and up to $\mathcal{O}(\as^3)$,
respectively; \nNNLO{} is the sum of $q\bar{q}$NNLO and $gg$NLO; 
\nNNLO{}$_{\rm bkg}$ is the corresponding 
cross section including only the continuum background without Higgs contributions,
whereas all other cross sections include resonant and non-resonant Higgs diagrams,
where applicable; \nNNLO{}$_{\rm EW}$ is our best prediction for the fiducial cross section.
It is obtained as in \citere{Grazzini:2020stb} for $WW$ production by including
EW corrections (to the $q\bar{q}$ channel) in a factorised approach \cite{Kallweit:2019zez}. 

With respect to the NLO cross section, the NNLO corrections in the $q\bar{q}$ channel
amount to $+6.3\%$ while the full NNLO corrections amount to $+15.1\%$.\footnote{Note that the 
NLO and NNLO $K$-factors in the \qqx{} channel are smaller here than in \citere{Grazzini:2018owa}
essentially due to the different choices of the PDFs at LO and NLO.}
Therefore, the loop-induced gluon fusion process contributes $58\%$ of the NNLO correction.
This is in line with previous computations \cite{Grazzini:2015hta,Grazzini:2018owa,Kallweit:2018nyv,Cascioli:2014yka,Grazzini:2017mhc}.
The \NLO{} corrections to the loop-induced contribution are huge, increasing 
$gg$LO by $+81.2\%$, which is even slightly higher than the $+70.8\%$ correction 
found with the setup considered in \citere{Grazzini:2018owa}, where the Higgs resonance region is excluded 
from the fiducial volume.
This confirms once more that those corrections depend on the fiducial cuts under consideration
and cannot be included through global rescaling factors.
The \nNNLO{} result is $+22.2\%$ higher than the NLO cross section, $+6.2$\%
higher than the NNLO cross section, and $+2.6\%$ higher than the \nNNLO{}$_{\rm bkg}$
Higgs background prediction. This means that the Higgs boson has a
positive contribution of $2.6\%$ at \nNNLO{}.
Finally, the EW corrections lead to a reduction of the cross section by $-5.6\%$,
which cancels almost exactly the contribution from the $gg$NLO corrections
so that the \nNNLO{}$_{\rm EW}$ prediction is only two permille above 
the NNLO cross section. However, while
such cancellation may occur at the level of the integrated cross section with a given set of fiducial cuts, $gg$NLO and 
NLO EW corrections have an effect in different regions of phase space and therefore 
do not compensate each other in differential distributions.

\begin{figure}[t]
\begin{center}                        
\begin{tabular}{c}
\includegraphics[width=.55\textwidth]{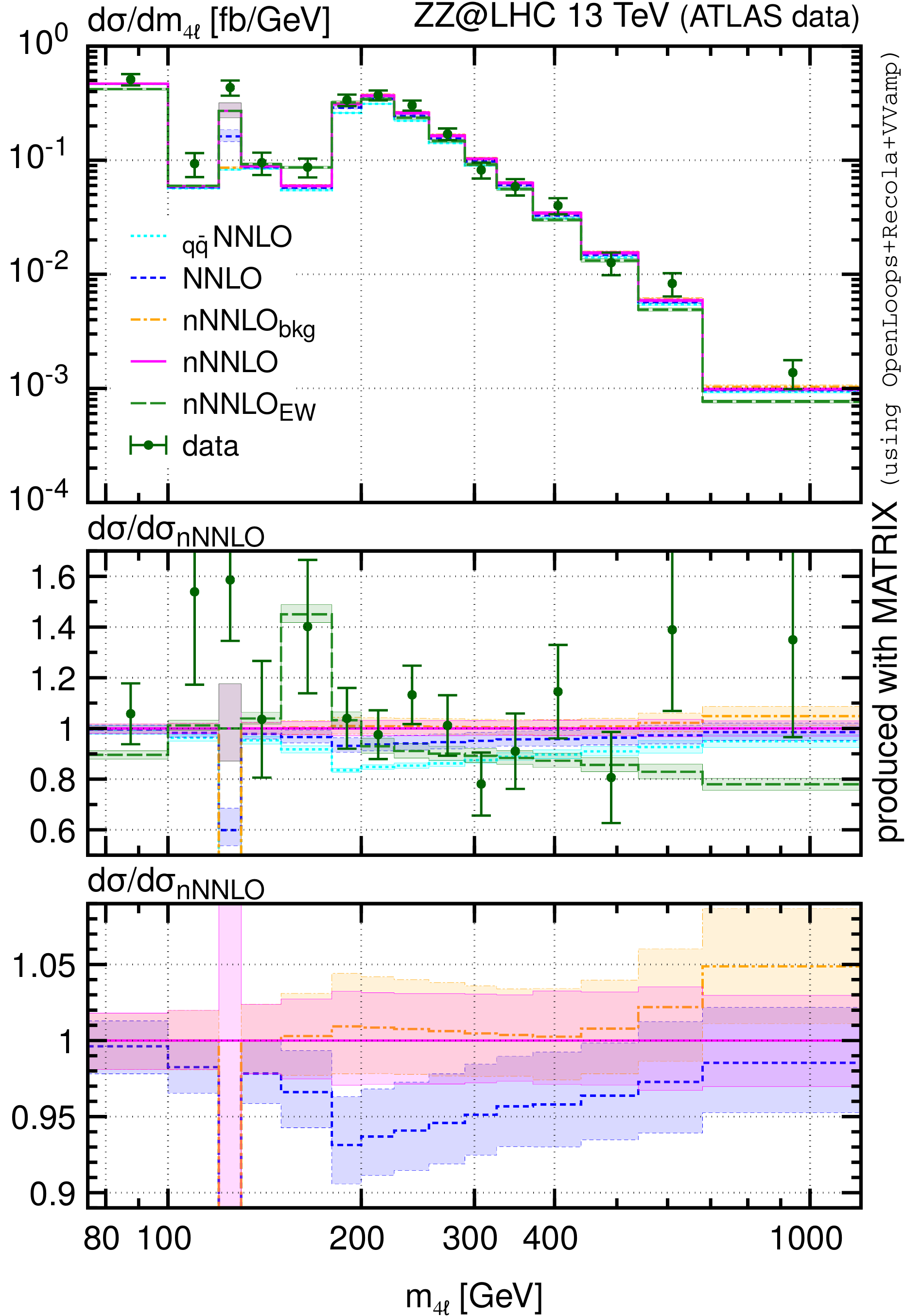} \\
\end{tabular}
\vspace*{2ex}
\caption{\label{fig:m_ZZ_4l} Invariant-mass distribution of the four leptons 
in the phase space volume defined in \citere{Aaboud:2019lxo} and summarized in \tab{tab:ZZcuts2}, compared to data from \citere{Aaboud:2019lxo}.}
\end{center}
\end{figure}

In \fig{fig:m_ZZ_4l} we show different predictions for the invariant-mass distribution
of the four leptons 
and compare them against ATLAS data from \citere{Aaboud:2019lxo} (green dots with experimental error bars).
In particular, we show $q{\bar q}$NNLO (light blue, dotted), \NNLO{} (blue, dashed), \nNNLO{} with (magenta, solid)
and without (orange, dash-dotted) Higgs contributions, and finally \nNNLO{} predictions including EW corrections (green, long-dashed).
The top panel shows the absolute distributions, while the lower two panels show predictions and data normalised to the \nNNLO{} result.
The agreement between theory and data is quite good. Unfortunately, 
the experimental uncertainties are still too large to clearly resolve the differences between
the various theoretical predictions. In particular, despite clear differences 
of the \nNNLO{} and \nNNLOEW{} predictions at high invariant mass and in the 
bin below the $2\,m_Z$ threshold, both predictions show a similar level 
of agreement to data, given the rather large experimental uncertainties. Nevertheless,
one can make the following two interesting observations: First, in that bin 
below the $2\,m_Z$ threshold, where large QED corrections are indeed 
expected, the \nNNLOEW{} prediction is in better agreement with the data point.
Second, in the tail of the invariant-mass distribution, where EW corrections have 
a large impact, data actually seem to be quite high and 
more consistent with the \nNNLO{} result.
Although this comparison has to be taken with caution due to the large experimental
errors in that region, this (small) $\sim 1.5\sigma$ excess over 
\nNNLOEW{} in the last two bins 
is an important demonstration of why EW corrections are so crucial:
If with decreasing experimental uncertainties one were to consider only the QCD 
prediction in such phase space region, an excess of the data over the actual SM 
prediction including EW corrections might go unnoticed.

We also find that in the region around $m_{4\ell}\sim 200$ GeV the NNLO prediction in the $q{\bar q}$ channel is almost $20\%$ smaller than the \nNNLO{} result,
which shows that also the loop-induced $gg$ channel yields a substantial contribution
to the cross section.
Indeed, the analysis of \citere{Aaboud:2019lxo} extracts a signal strength of the loop-induced gluon fusion contribution of $\mu_{gg}=1.3\pm 0.5$.
Note that the $gg$ contribution becomes even larger in the bin around 
$m_{4\ell}=125$\,GeV due to the Higgs resonance. 
In that bin $q\bar{q}$NNLO and \nNNLO{}$_{\rm bkg}$ 
predictions are way below data, since they do not include 
resonant Higgs contributions. Also the NNLO prediction is quite low, since it 
misses the large NLO corrections to Higgs production. However, one should bear
in mind that also the full \nNNLO{} prediction misses the relatively large 
higher-order corrections to on-shell Higgs production beyond NLO (see \citere{Anastasiou:2016cez} and references therein).

In the bottom panel we increase the resolution of the relative 
differences to \nNNLO{} for a subset of the QCD predictions. Comparing the \NNLO{} and \nNNLO{} results we see that their uncertainty bands overlap
almost everywhere. The largest effect of \nNNLO{} corrections is in the region $m_{4\ell}\sim 200$ GeV where the difference with \NNLO{} is about $7\%$.
We also notice that the Higgs background prediction departs from the full result as $m_{4\ell}$ increases, where it becomes larger. The effect is about $+5\%$ 
in the last $m_{4\ell}$ bin.
This means that in this region the relative impact of the Higgs contribution is negative and becomes increasingly large, which is caused by the Higgs signal--background interference.
In the following, we will investigate in more detail the relative effects when 
separating Higgs signal, background and interference contributions.

\input{tables/XS_ZZ_ggNLO_tail_resonance.tex}

We now continue our presentation of phenomenological results 
by studying the theoretical ingredients used in Higgs off-shell studies to constrain $\Gamma_H$
at the LHC. The relevant quantity is the ratio of the off-shell to the on-shell Higgs cross section \cite{Aaboud:2018puo,Sirunyan:2019twz}.
To this end, we report in \refta{tab:XS_ZZ_m4l_ranges} various contributions to the fiducial cross section in the off-shell region
with $m_{4\ell}>200$ GeV (left) and in the Higgs signal region \mbox{120~GeV $<m_{4\ell}<$130~GeV}~(right).
Besides the notation already introduced in the discussion of \tab{tab:XS_ZZ_2}, we use the 
abbreviations ``sig", ``bkg", and ``intf" to separate 
the $4\ell$ Higgs signal contribution, 
the $4\ell$ continuum background contribution, and their interference, respectively.
We recall that this separation is needed when constraining the Higgs 
width at the LHC \cite{Aaboud:2018puo,Sirunyan:2019twz}. In particular, in the scenario 
proposed in \citere{Caola:2013yja} the Higgs couplings and width are rescaled 
from their SM values such that the on-shell cross section remains unchanged. 
In this case, the off-shell Higgs rate needs to be evaluated by adding the Higgs signal contribution rescaled by 
$\Gamma_H/\Gamma_H^{\rm SM}$ and the Higgs interference contribution rescaled by $\sqrt{\Gamma_H/\Gamma_H^{\rm SM}}$. 
Accurate predictions of the separate contributions of the Higgs boson signal, the background and their interference are therefore 
indispensible for such analyses.

We start our discussion from the region $m_{4\ell}>200$ GeV. We see that in this region the interference is negative, as expected from unitarity arguments.
Therefore, the gluon fusion cross section is smaller than the sum of the signal and background cross sections by about $11\%$ both at LO and at NLO.
In particular, the interference is almost twice as large as the signal in absolute value, and its size is about $12\%$ compared to the background,
which in turn is only about $17\%$ of the NNLO result in the $q{\bar q}$ channel.
The large cancellations between signal and interference render the separation of the off-shell Higgs cross section from the background difficult.
We note that, as argued in early off-shell studies (see e.g. Ref.~\cite{Heinemeyer:2013tqa}),
the NLO $K$-factor for the interference is very close to the geometrical average of the $K$-factors for signal and background.
However, we stress that this conclusion is strongly dependent on the fiducial cuts and setup under consideration.

We now continue our discussion of \tab{tab:XS_ZZ_m4l_ranges} with the region 120~GeV $<m_{4\ell}<$130~GeV.
As expected, the Higgs signal cross section is by far dominant due to resonant Higgs contributions, being about 
60 times larger than the gluon fusion background.
The interference is positive, but about two orders of magnitude smaller than the background.
It is worth noticing that the size of the NLO corrections for signal, background and interference is relatively similar 
when the Higgs boson is off-shell, whereas in the region where the Higgs boson can become on-shell the NLO $K$-factor of the 
signal contribution is significantly larger than that of background and interference.

\begin{figure}[p]
\begin{center}                        
\begin{tabular}{cc}
\includegraphics[width=.48\textwidth]{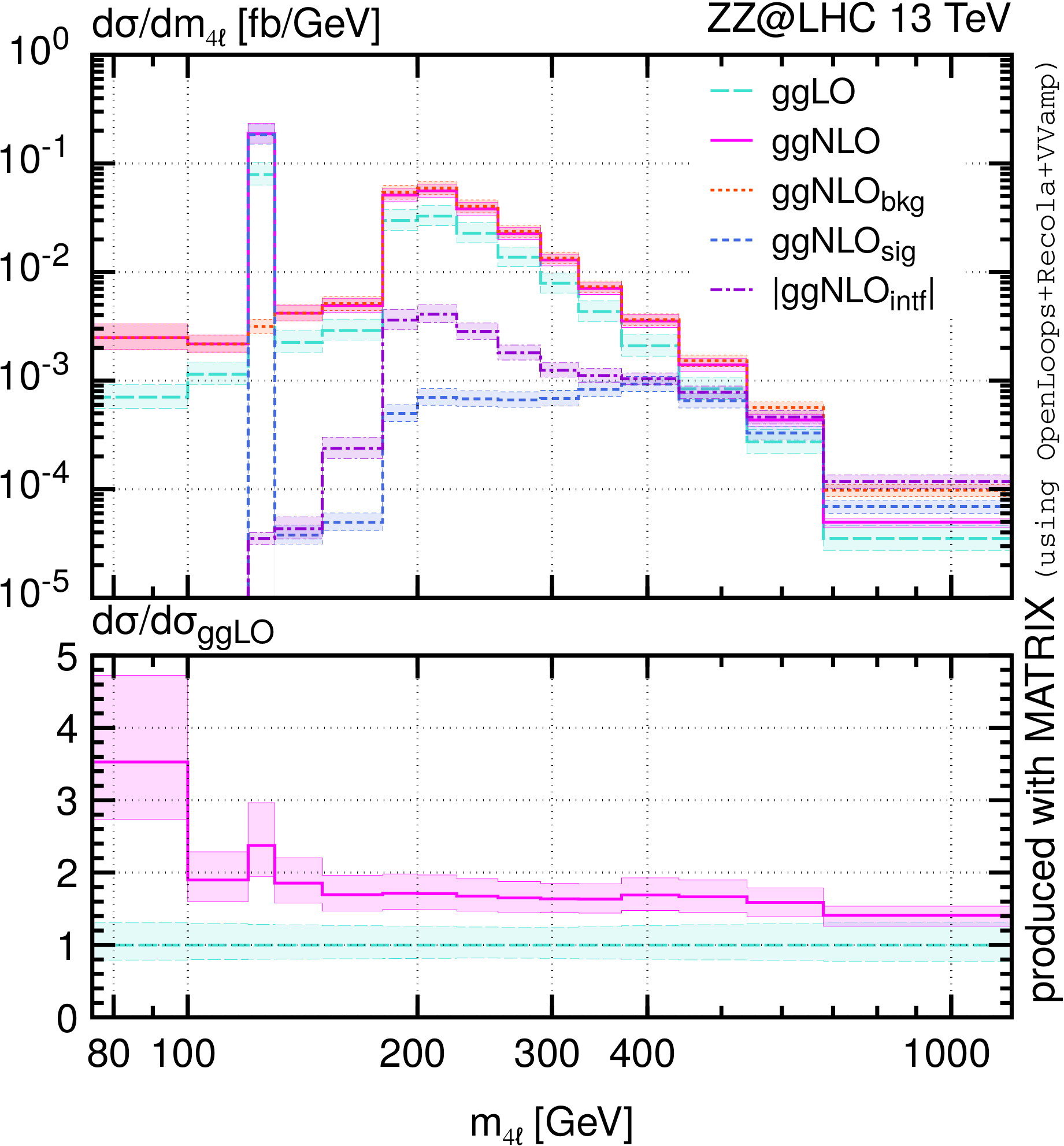} &
\includegraphics[width=.48\textwidth]{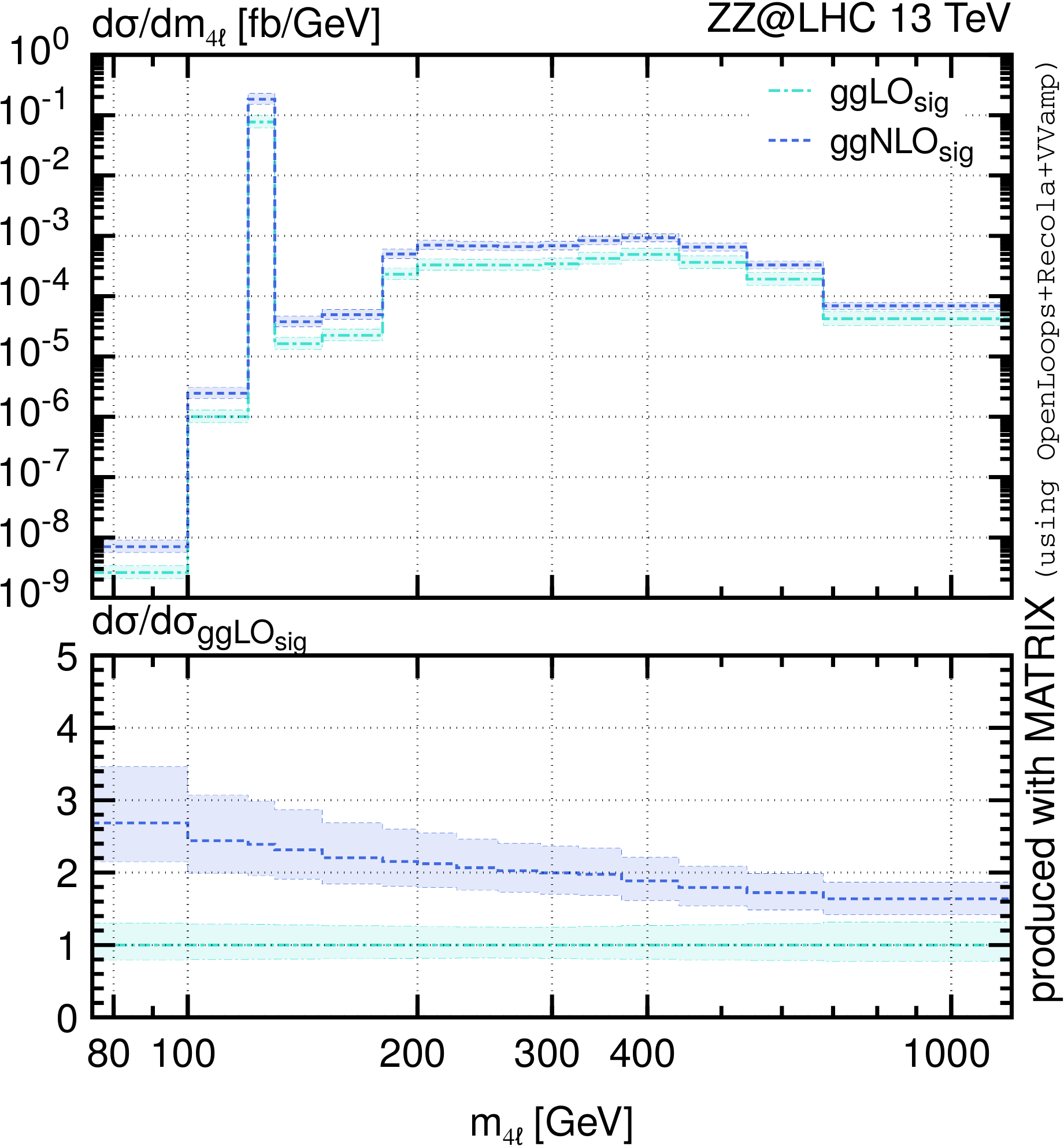} \\[1ex]
\quad (a) & \quad (b)\\[2ex]
\includegraphics[width=.48\textwidth]{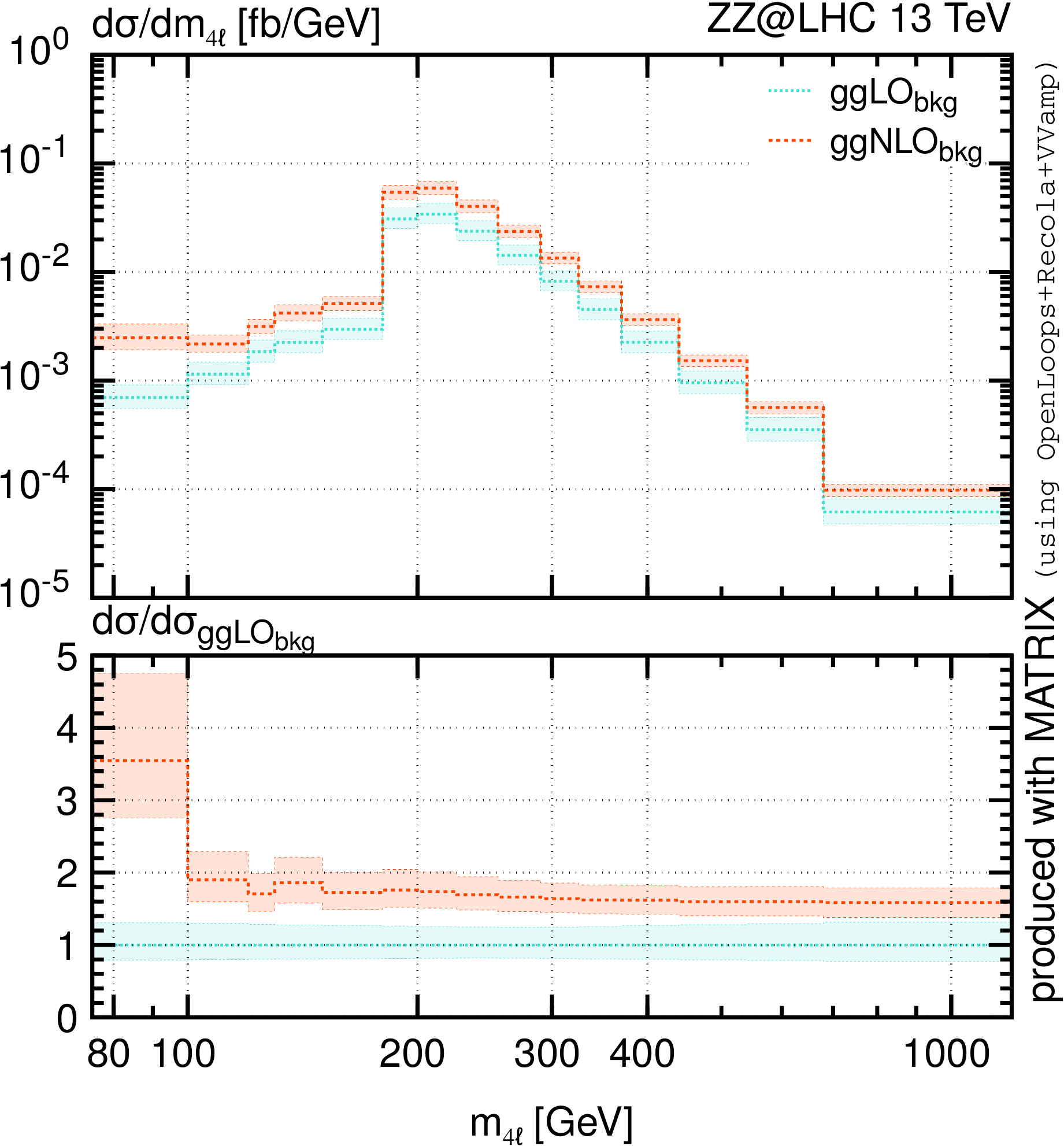} &
\includegraphics[width=.48\textwidth]{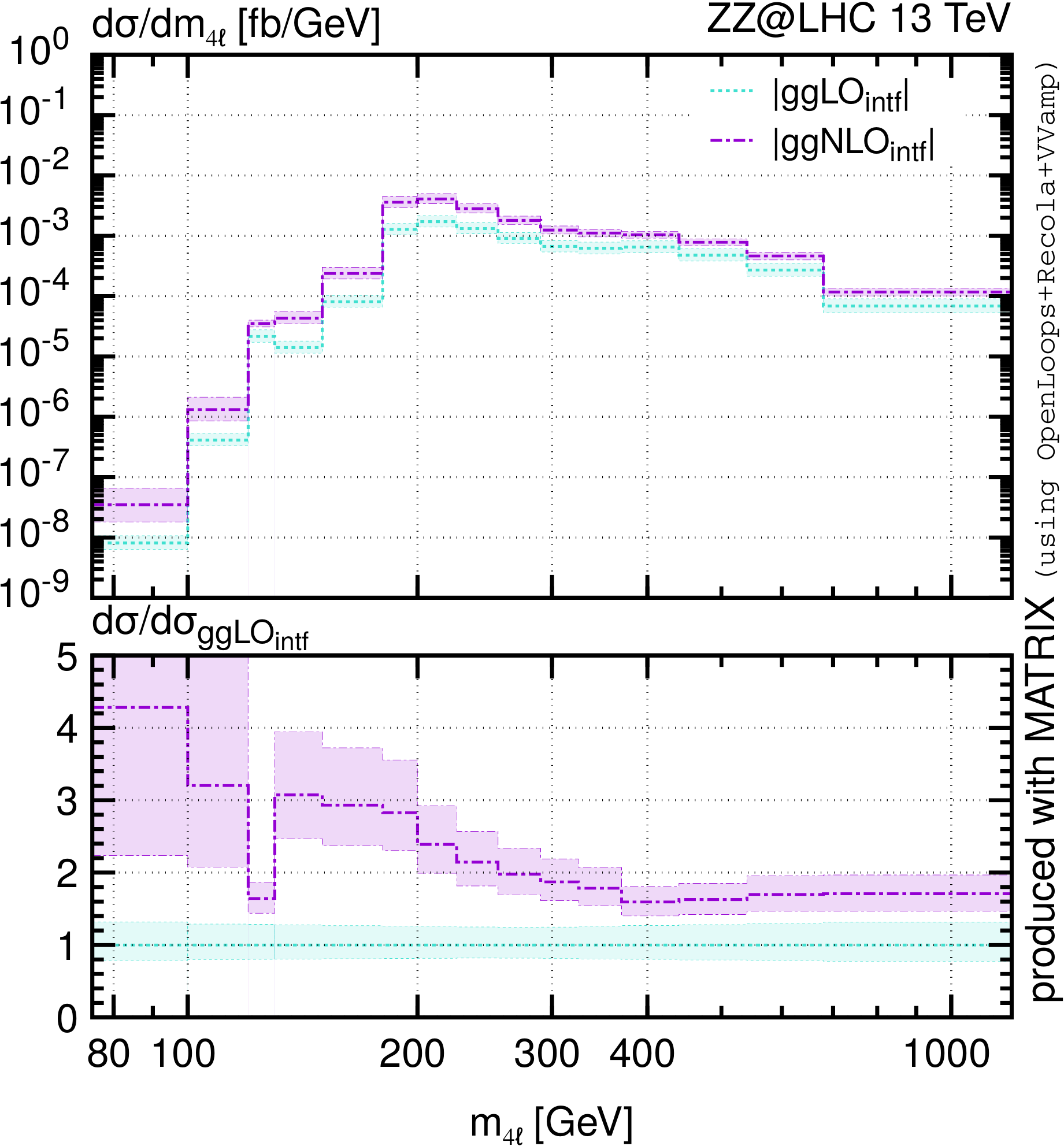} \\[1ex]
\quad (c) & \quad (d)\\[2ex]
\end{tabular}
\caption{\label{fig:m_ZZ_4l_NLOgg} Four-lepton invariant mass distribution for the loop-induced $gg$ channel with the phase space definition of \citere{Aaboud:2019lxo}.
  We show the full result (a), as well as the contributions of the Higgs signal (b), background (c) and interference (d) separately.}
\end{center}
\end{figure}

In \reffi{fig:m_ZZ_4l_NLOgg} we study the behaviour of the signal, background and interference contributions
to the invariant-mass distribution in the loop-induced gluon fusion channel. We use the same invariant-mass range and binning as 
considered in the ATLAS analysis \cite{Aaboud:2019lxo}.
In \reffi{fig:m_ZZ_4l_NLOgg}\,(a) we show the full result at LO (turquoise, long-dashed) and NLO (magenta, solid). For comparison, 
also the signal (blue, dashed), the background (red, dotted), and the modulus of the interference contribution (purple, dash-dotted) are shown at NLO in the main frame.
The separate LO and NLO results for the signal, the background and the interference are presented in \reffi{fig:m_ZZ_4l_NLOgg}\,(b), (c), and (d), respectively.
In the lower panels we study the different behaviour of the NLO $K$-factors, i.e. the ratios of the NLO to the LO predictions.

For the signal contribution we clearly see the peak at $m_{4\ell}=125$ GeV from the Higgs resonance, and then the cross section quickly drops and increases again at the $2m_Z$ threshold,
remaining roughly constant up about $400$\,GeV where it starts to decrease again.
Above $400$\,GeV the signal and the interference are of the same order,
while for $200\lesssim m_{4\ell}\lesssim 400$\,GeV the absolute value of the negative interference contribution is even larger.
These features are well known \cite{Kauer:2012hd,Passarino:2013bha}: the decrease of the signal cross section due to the off-shell Higgs boson propagator
is compensated by the $|{\cal A}|^2\sim m^4_{4\ell}$ increase of the decay amplitude, thereby leading to the plateau observed in  \reffi{fig:m_ZZ_4l_NLOgg} (b).
For the signal the impact of the NLO corrections is about $+170\%$ at small invariant masses,
and it slowly decreases as $m_{4\ell}$ increases, being about $+60\%$ in the high-mass region.
The background distribution has a broad maximum for $m_{4\ell}\gtrsim 2m_Z$ due to the $ZZ$ resonance, while the impact of NLO corrections is more uniform,
ranging from about $100\%$ in the second bin to $60\%$ in the high-$m_{4\ell}$ region.
The interference is negative and peaked at $m_{4\ell}\sim 200$ GeV, but it changes sign in the Higgs signal region.
In the region $m_{4\ell}\sim 200$ GeV the NLO corrections to the interference are very large (about $+150\%$), and they are larger than for the signal and the background,
decreasing to about $70\%$ at large values of $m_{4\ell}$.

In conclusion, in all cases radiative corrections have the effect 
of increasing the absolute size of the individual contributions. 
However, the relative size of the corrections 
for the individual contributions is quite different, especially at 
small $m_{4\ell}$ values, and the full result is a combination 
of all of those effects. Only at large invariant masses ($m_{4\ell}\gtrsim 400$\,GeV) the 
relative size of the corrections becomes similar for signal, background and
interference. It is therefore difficult to make a direct connection
between the QCD corrections beyond NLO for the signal, 
which are known to be relatively large (see \citere{Anastasiou:2016cez} and references therein), and the other contributions,
where they are not known.
Nevertheless, the NLO corrections in the off-shell region 
are not that different among the three contributions, and the QCD effects beyond NLO are expected to 
be significant. Therefore, in order to approximately take higher-order corrections into account, one might be tempted to rescale 
our NLO result for the off-shell cross section by using the relative impact of the QCD corrections
beyond NLO evaluated in the off-shell region for the signal contribution \cite{Anastasiou:2016cez}.
Needless to say, much care should be taken when following such approach.

In this Letter, we have studied the production of four charged leptons in $pp$ collisions at 13 TeV,
and we have computed the NLO QCD corrections to the loop-induced gluon fusion contribution.
Our computation consistently accounts for the Higgs boson signal, its corresponding background and their interference.
The contribution from heavy-quark loops is exactly included in the calculation except for the two-loop $gg\to ZZ \to 4\ell$ diagrams,
for which the heavy-quark effects are approximated through a reweighting procedure.
Our calculation is combined with the NNLO QCD and NLO EW corrections in the quark-annihilation channel, and it includes all partonic channels,
spin correlations and off-shell effects.
The computation is implemented in the \Matrix{} framework and allows us to separately study the Higgs boson signal,
the background and the interference contributions. Those are the central theoretical 
ingredients of experimental analyses that place bounds on the total
Higgs boson width. In particular, for the background and the interference our calculation constitutes the most advanced prediction.
We look forward to applications of this calculation and the corresponding implementation in \Matrix{} to off-shell Higgs boson studies at the LHC and beyond.

\noindent {\bf Acknowledgements.}
We thank Fabrizio Caola and Raoul R\"{o}ntsch
for discussions and for providing details on their computations and results.
We would also like to express our gratitude to Jean-Nicholas Lang, Jonas Lindert and Federico Buccioni for providing private amplitudes and clarifications.
This work is supported in part by the Swiss National Science Foundation (SNF) under contract 200020$\_$188464.
The work of JY is supported by Forschungskredit der Universit\"{a}t Z\"{u}rich, Verf\"{u}gung Nr. [FK-19-092], and that of SK by the ERC Starting Grant 714788 REINVENT.


\bibliographystyle{apsrev4-1}
\bibliography{zznlogg_off}
\end{document}

%% file: defs.tex
\providecommand{\href}[2]{#2}

\newcommand\as{\alpha_{\mathrm{S}}}

\def\to{\rightarrow}

\newcommand{\GeV}{\ensuremath{\rm GeV}}
\newcommand{\TeV}{\ensuremath{\rm TeV}}

\newcommand\Matrix{{\sc Matrix}\xspace}
\newcommand\Munich{{\sc Munich}\xspace}
\newcommand\OpenLoops{{\sc OpenLoops}\xspace}

\newcommand\Recola{{\sc Recola}\xspace}

\newcommand{\eqn}[1]{Eq.\,(\ref{#1})}

\newcommand{\fig}[1]{Figure~\ref{#1}}

\newcommand{\tab}[1]{Table~\ref{#1}}

\def\reffi#1{\mbox{Figure~\ref{#1}}}

\def\refta#1{\mbox{Table~\ref{#1}}}

\newcommand{\citere}[1]{\mbox{Ref.~\cite{#1}}}
\newcommand{\citeres}[1]{\mbox{Refs.~\cite{#1}}}

\renewcommand{\gg}{\ensuremath{gg}\xspace} 
\newcommand{\qqx}{\ensuremath{q\bar{q}}\xspace}
\newcommand{\qg}{\ensuremath{qg}\xspace}

\newcommand{\zz}{\ensuremath{ZZ}\xspace}

\newcommand{\abbrev}{}

\newcommand{\LO}{\ensuremath{\rm{\abbrev LO}}\xspace}
\newcommand{\lo}{\LO}
\newcommand{\NLO}{\ensuremath{\rm{\abbrev NLO}}\xspace}
\newcommand{\nlo}{\NLO}
\newcommand{\NNLO}{\ensuremath{\rm{\abbrev NNLO}}\xspace}

\newcommand{\qqNNLO}{\ensuremath{q\bar{q}{\rm{\abbrev NNLO}}}\xspace}
\newcommand{\ggLO}{\ensuremath{gg{\rm{\abbrev LO}}}\xspace}
\newcommand{\ggNLO}{\ensuremath{gg{\rm{\abbrev NLO}}}\xspace}

\newcommand{\nNNLO}{\ensuremath{{\rm n{\abbrev NNLO}}}\xspace}
\newcommand{\nNNLOEW}{\ensuremath{{\rm n{\abbrev NNLO}_{EW}}}\xspace}

\newcommand{\mfourl}{\ensuremath{m_{4\ell}}}

%% file: tables/Cuts_ATLAS.tex
\setlength{\tabcolsep}{10pt}
\renewcommand{\arraystretch}{1.5}
\begin{table}[t]
\begin{center}
\begin{tabular}{|c|}
\hline
definition of the fiducial volume for $pp\to 4 \ell +X$\\
\hline
muon selection with $p_{T,\mu}>5$\,GeV and $|\eta_\mu| < 2.7$ \\
electron selection with $p_{T,e}>7$\,GeV and $|\eta_e| < 2.47$\\
$p_{T, \ell_1} > 20$ GeV, $p_{T, \ell_2} > 15$ GeV and $p_{T, \ell_3} > 10$ GeV for leading three leptons\\
$50 \, \GeV{} < m_{12} < 106 \, \GeV{}$ and $f(m_{4 \ell}) < m_{34} < 115 \, \GeV{}$\\
$\Delta R_{\ell_i \ell_j} >0.1(0.2)$ for same (opposite) flavour leptons \\
$m_{\ell_i \ell_j} > 5$ GeV for all same-flavour opposite-sign pairs \\
$70\,\GeV{} < m_{4 \ell} < 1200$ GeV\\
\hline
\end{tabular}
\end{center}
\caption{\label{tab:ZZcuts2} 
Fiducial phase-space definitions of the ATLAS \zz{} measurements at $\sqrt{s} = 13$\,TeV~\cite{Aaboud:2019lxo}.}
\end{table}
\renewcommand{\arraystretch}{1.0}
\setlength{\tabcolsep}{5pt}

%% file: tables/XS_ZZ_NNPDF31luxqedNf4.tex
\renewcommand\arraystretch{1.8}
\begin{table}[t]
\begin{center}
\begin{tabular}{|c|c|c|}
\hline
$\sqrt{s}=13$ TeV & $\sigma$\,[fb] & $\sigma/\sigma_{\rm NLO}-1$ \\
\hline
\lo{}       &\hspace*{.2em} $36.848(1)\phantom{0}\,^{\phantom{0}+7.1\%}_{\phantom{0}-8.1\%}$ \hspace*{.2em}& $-24.8\%$ \\
\nlo{}      &\hspace*{.2em} $48.990(2)\phantom{0}\,^{\phantom{0}+3.1\%}_{\phantom{0}-2.9\%}$ \hspace*{.2em}& --- \\
\qqNNLO{}   &\hspace*{.2em} $52.07(4)\phantom{00}\,^{\phantom{0}+1.4\%}_{\phantom{0}-1.4\%}$ \hspace*{.2em}&  $\phantom{0}$$+6.3\%$ \\
\hline
 & $\sigma$\,[fb] & $\sigma/\sigma_{\rm{ggLO}}-1$ \\
\hline
\ggLO{}     &\hspace*{.2em} $\phantom{0}4.2967(3)\,^{+25.6\%}_{-18.4\%}$ \hspace*{.2em}&  ---  \\
\ggNLO{}    &\hspace*{.2em} $\phantom{0}7.80(2)\phantom{00}\,^{+17.1\%}_{-13.9\%}$ \hspace*{.2em}&  $+81.5\%$  \\
\hline
& $\sigma$\,[fb] & $\sigma/\sigma_{\rm{NLO}}-1$ \\
\hline
\NNLO{}     &\hspace*{.2em} $56.37(4)\phantom{00}\,^{\phantom{0}+3.2\%}_{\phantom{0}-2.7\%}$ \hspace*{.2em}&  $+15.1\%$ \\
\nNNLO{}    &\hspace*{.2em} $59.87(4)\phantom{00}\,^{\phantom{0}+3.4\%}_{\phantom{0}-3.1\%}$ \hspace*{.2em}& $+22.2\%$ \\
\nNNLO{}$_{\rm bkg}$ &\hspace*{.2em} $58.37(4)\phantom{00}\,^{\phantom{0}+2.8\%}_{\phantom{0}-2.6\%}$ \hspace*{.2em}& $+19.1\%$ \\
\hline
 & $\sigma$\,[fb] & $\sigma/\sigma_{\rm{nNNLO}}-1$ \\
\hline
\nNNLOEW{}  &\hspace*{.2em} $56.49(4)\phantom{00}\,^{\phantom{0}+3.5\%}_{\phantom{0}-3.1\%}$ \hspace*{.2em}& $\phantom{0}$$-5.6\%$  \\
\hline
\end{tabular}
\end{center}
\caption{\label{tab:XS_ZZ_2} Fiducial cross sections in the phase space volume defined in \citere{Aaboud:2019lxo} and summarized in \tab{tab:ZZcuts2} at different perturbative orders. Statistical uncertainties for (n)NNLO results include the uncertainties due the $r_{\mathrm{cut}}$ extrapolation in $q_T$ subtraction \cite{Grazzini:2017mhc}.}
\end{table}
\renewcommand{\arraystretch}{1.0}

%% file: tables/XS_ZZ_ggNLO_tail_resonance.tex
\renewcommand{\arraystretch}{1.8}
\begin{table}[p]
\begin{center}
\begin{tabular}{|c|cc|cc|}\hline
$\sqrt{s}=13\,\mathrm{TeV}$ & \multicolumn{2}{c|}{$m_{4\ell} > 200 \,\mathrm{GeV}$} & \multicolumn{2}{c|}{$120 < m_{4\ell} < 130 \,\mathrm{GeV} $}  \\
\hline
& $\sigma\, (\mathrm{fb})$ & ${\sigma}/{\sigma_{gg\mathrm{LO}}}-1$ & $\sigma\, (\mathrm{fb})$ & ${\sigma}/{\sigma_{gg\mathrm{LO}}}-1$ \\
\hline
$gg\mathrm{LO}$ &     $   \phantom{-0}2.73726(28)\phantom{0}\,^{   +25.32 \%}_{   -18.57 \%}$ & --- &    \hspace*{.4em} $   0.78952(12)\phantom{0}\,^{   +28.48 \%}_{   -19.83 \%}$ & --- \\
$gg\mathrm{NLO}$ &     $    \phantom{-0}4.5790(53)\phantom{00}\,^{   +14.24 \%}_{   -12.46 \%}$ & $    +67.3 \%$ &    \hspace*{.4em} $    1.8745(87)\phantom{00}\,^{   +24.88 \%}_{   -18.02 \%}$ & $   +137.4 \%$ \\
\hline
& $\sigma\, (\mathrm{fb})$ & ${\sigma}/{\sigma_{gg\mathrm{LO}}^{\mathrm{bkg}}}-1$ & $\sigma\, (\mathrm{fb})$ & ${\sigma}/{\sigma_{gg\mathrm{LO}}^{\mathrm{bkg}}}-1$ \\
\hline
$gg\mathrm{LO}_{\mathrm{bkg}}$ &     \hspace*{-.5em} $\phantom{-0}2.89117(27)\phantom{0}\,^{   +25.38 \%}_{   -18.61 \%}$ & --- &    \hspace*{.4em} $  0.018466(31)\,^{   +28.47 \%}_{   -19.83 \%}$ & --- \\
$gg\mathrm{NLO}_{\mathrm{bkg}}$ &    \hspace*{-.5em} $\phantom{-0}4.8615(33)\phantom{00}\,^{   +14.32 \%}_{   -12.48 \%}$ & $    +68.2 \%$ &    \hspace*{.4em} $    0.0315(32)\phantom{00}\,^{   +16.61 \%}_{   -13.96 \%}$ & $\phantom{0}$$    +70.6 \%$ \\
\hline
& $\sigma\, (\mathrm{fb})$ & $\sigma/\sigma_{gg\mathrm{LO}}^{\mathrm{intf}}-1$ & $\sigma\, (\mathrm{fb})$ & $\sigma/\sigma_{gg\mathrm{LO}}^{\mathrm{intf}}-1$ \\
\hline
$gg\mathrm{LO}_{\mathrm{intf}}$ &  \hspace*{-.5em} $\phantom{0}$$-0.333378(29)\,^{   -19.76 \%}_{   +26.97 \%}$ & --- &     \hspace*{.4em} $  0.000215(10)\,^{   +28.55 \%}_{   -19.86 \%}$ & --- \\
$gg\mathrm{NLO}_{\mathrm{intf}}$ &    \hspace*{-.5em} $\phantom{0}$$-0.6174(42)\phantom{00}\,^{   -14.35 \%}_{   +17.18 \%}$ & $    +85.2 \%$ &    \hspace*{.4em} $   0.00035(33)\phantom{0}\,^{   +13.46 \%}_{   -12.40 \%}$ & $\phantom{0}$$    +64.3 \%$ \\
\hline
& $\sigma\, (\mathrm{fb})$ & ${\sigma}/{\sigma_{gg\mathrm{LO}}^{\mathrm{sig}}}-1$ & $\sigma\, (\mathrm{fb})$ & ${\sigma}/{\sigma_{gg\mathrm{LO}}^{\mathrm{sig}}}-1$ \\
\hline
$gg\mathrm{LO}_{\mathrm{sig}}$ &    \hspace*{-.5em} $\phantom{-0}0.180110(16)\,^{   +27.50 \%}_{   -20.19 \%}$ & --- &    \hspace*{.4em} $  0.770793(29)\,^{   +28.48 \%}_{   -19.83 \%}$ & --- \\
$gg\mathrm{NLO}_{\mathrm{sig}}$ &    \hspace*{-.5em} $\phantom{-0}0.33555(14)\phantom{0}\,^{   +17.15 \%}_{   -14.35 \%}$ & $    +86.3 \%$ &    \hspace*{.4em} $    1.8426(81)\phantom{00}\,^{   +25.03 \%}_{   -18.09 \%}$ & $   +139.1 \%$ \\
\hline
& $\sigma\, (\mathrm{fb})$ & ${\sigma}/{\sigma_{\mathrm{NLO}}}-1$ & $\sigma\, (\mathrm{fb})$ & ${\sigma}/{\sigma_{\mathrm{NLO}}}-1$ \\
\hline
$\mathrm{LO}$ &    \hspace*{-.5em} $\phantom{-}21.37744(43)\phantom{0}\,^{\phantom{0}    +4.15 \%}_{\phantom{0}    -5.08 \%}$ & $    -23.7 \%$ &    \hspace*{.4em} $  0.633439(77)\,^{   +11.31 \%}_{   -12.46 \%}$ & $\phantom{0}$$    -19.8 \%$ \\
$\mathrm{NLO}$ &    \hspace*{-.5em} $\phantom{-}28.02236(77)\phantom{0}\,^{ \phantom{0}   +2.80 \%}_{\phantom{0}    -2.31 \%}$ & --- &    \hspace*{.4em} $   0.78944(16)\phantom{0}\,^{\phantom{0}    +2.84 \%}_{\phantom{0}    -4.64 \%}$ & --- \\
$q\bar{q}\mathrm{NNLO}$ &    \hspace*{-.5em} $\phantom{-}29.887(12)\phantom{000}\,^{\phantom{0}    +1.45 \%}_{\phantom{0}    -1.41 \%}$ & $\phantom{0}$$     +6.7 \%$ &    \hspace*{.4em} $    0.8296(26)\phantom{00}\,^{\phantom{0}    +1.15 \%}_{\phantom{0} -1.22 \%}$ & $\phantom{00}$$     +5.1 \%$ \\
$\mathrm{NNLO}$ &    \hspace*{-.5em} $\phantom{-}32.625(12)\phantom{000}\,^{\phantom{0}    +3.44 \%}_{\phantom{0}    -2.83 \%}$ & $    +16.4 \%$ &    \hspace*{.4em} $    1.6191(26)\phantom{00}\,^{   +14.48 \%}_{   -10.30 \%}$ & $   +105.1 \%$ \\
$\mathrm{nNNLO}$ &    \hspace*{-.5em} $\phantom{-}34.466(13)\phantom{000}\,^{\phantom{0}    +3.13 \%}_{\phantom{0}    -2.87 \%}$ & $    +23.0 \%$ &    \hspace*{.4em} $    2.7041(91)\phantom{00}\,^{   +17.60 \%}_{   -12.87 \%}$ & $   +242.5 \%$ \\
$\mathrm{nNNLO_{EW}}$ &    \hspace*{-.5em} $\phantom{-}31.052(12)\phantom{000}\,^{\phantom{0}    +3.31 \%}_{\phantom{0}    -3.02 \%}$ & $    +10.8 \%$ &    \hspace*{.4em} $    2.7043(91)\phantom{00}\,^{   +17.60 \%}_{   -12.87 \%}$ & $   +242.6 \%$ \\
\hline
\end{tabular}
\end{center}
\caption{\label{tab:XS_ZZ_m4l_ranges}
  Integrated cross sections in the four-lepton invariant-mass ranges
  $m_{4\ell} > 200 \,\mathrm{GeV}$
  and $120 < m_{4\ell} < 130 \,\mathrm{GeV} $ in the fiducial phase space defined in \citere{Aaboud:2019lxo}, at different perturbative orders.
  The Higgs signal (sig), background (bkg), and interference (intf) contributions in the loop-induced $gg$ channel are stated separately.}
\end{table}
\renewcommand{\arraystretch}{1.0}

%% file: zznlogg_off.bbl
\begin{thebibliography}{94}
\bibitem [{mun()}]{munich}%
  \BibitemOpen
  \href@noop {} {}\bibinfo {note} {{\sc Munich} is the abbreviation of
  ``MUlti-chaNnel Integrator at Swiss (CH) precision''---an automated parton
  level NLO generator by S.~Kallweit. In preparation.}\BibitemShut {Stop}%
\end{thebibliography}
